\input harvmac
\noblackbox
\input epsf.tex
\def\N{{\cal N}}
\def\O{{\cal O}}

\def\F{{\cal F}}
\def\V{{\rm Vol}}
\def\L{L_{\alpha \beta}}

\lref\SG{
S.~S.~Gubser,
``Einstein manifolds and conformal field theories,''
Phys.\ Rev.\ D {\bf 59}, 025006 (1999)
[arXiv:hep-th/9807164].
}

\lref\DMRP{
D.~R.~Morrison and M.~R.~Plesser,
``Non-spherical horizons. I,''
Adv.\ Theor.\ Math.\ Phys.\  {\bf 3}, 1 (1999)
[arXiv:hep-th/9810201].
}

\lref\HV{
K.~Hori and C.~Vafa,
``Mirror symmetry,''
arXiv:hep-th/0002222.
}

\lref\HIV{
K.~Hori, A.~Iqbal and C.~Vafa,
``D-branes and mirror symmetry,''
arXiv:hep-th/0005247.
}

\lref\CFIKV{
F.~Cachazo, B.~Fiol, K.~A.~Intriligator, S.~Katz and C.~Vafa,
``A geometric unification of dualities,''
Nucl.\ Phys.\ B {\bf 628}, 3 (2002)
[arXiv:hep-th/0110028].
}

\lref\BH{
A.~Bergman and C.~P.~Herzog,
``The volume of some non-spherical horizons and the AdS/CFT  correspondence,''
JHEP {\bf 0201}, 030 (2002)
[arXiv:hep-th/0108020].
}

\lref\CBRP{
C.~E.~Beasley and M.~R.~Plesser,
``Toric duality is Seiberg duality,''
JHEP {\bf 0112}, 001 (2001)
[arXiv:hep-th/0109053].
}

\lref\SeibergPQ{
N.~Seiberg,
``Electric - magnetic duality in supersymmetric nonAbelian gauge theories,''
Nucl.\ Phys.\ B {\bf 435}, 129 (1995)
[arXiv:hep-th/9411149].
}

\lref\Wbary{
E.~Witten,
``Baryons and branes in anti de Sitter space,''
JHEP {\bf 9807}, 006 (1998)
[arXiv:hep-th/9805112].
}

\lref\IH{
A.~Hanany and A.~Iqbal,
``Quiver theories from D6-branes via mirror symmetry,''
JHEP {\bf 0204}, 009 (2002)
[arXiv:hep-th/0108137].
}

\lref\BHK{
D.~Berenstein, C.~P.~Herzog and I.~R.~Klebanov,
``Baryon spectra and AdS/CFT correspondence,''
JHEP {\bf 0206}, 047 (2002)
[arXiv:hep-th/0202150].
}

\lref\BG{
C.~P.~Boyer and K.~Galicki,
``On Sasakian-Einstein Geometry,''
Int.\ J.\ Math.\  {\bf 11}, 873 (2000)
[arXiv:math.dg/9811098].
}

\lref\WH{
E.~Witten,
``Anti-de Sitter space and holography,''
Adv.\ Theor.\ Math.\ Phys.\  {\bf 2}, 253 (1998)
[arXiv:hep-th/9802150].
}

\lref\DM{
M.~R.~Douglas and G.~W.~Moore,
``D-branes, Quivers, and ALE Instantons,''
arXiv:hep-th/9603167.
}

\lref\LNV{
A.~E.~Lawrence, N.~Nekrasov and C.~Vafa,
``On conformal field theories in four dimensions,''
Nucl.\ Phys.\ B {\bf 533}, 199 (1998)
[arXiv:hep-th/9803015].
}

\lref\FHHI{
B.~Feng, A.~Hanany, Y.~H.~He and A.~Iqbal,
``Quiver theories, soliton spectra and Picard-Lefschetz transformations,''
JHEP {\bf 0302}, 056 (2003)
[arXiv:hep-th/0206152].
}

\lref\FFHH{
B.~Feng, S.~Franco, A.~Hanany and Y.~H.~He,
``Unhiggsing the del Pezzo,''
arXiv:hep-th/0209228.
}

\lref\MW{
M.~Wijnholt,
``Large volume perspective on branes at singularities,''
arXiv:hep-th/0212021.
}

\lref\AEFJ{D.~Anselmi, J.~Erlich, D.~Z.~Freedman and A.~A.~Johansen,
``Positivity constraints on anomalies in supersymmetric gauge theories,''
Phys.\ Rev.\ D {\bf 57}, 7570 (1998)
[arXiv:hep-th/9711035].
}

\lref\AFGJ{D.~Anselmi, D.~Z.~Freedman, M.~T.~Grisaru and A.~A.~Johansen,
``Nonperturbative formulas for central functions of supersymmetric
gauge theories,'' Nucl.\ Phys.\ B {\bf 526}, 543 (1998)
[arXiv:hep-th/9708042].
}

\lref\WittenQJ{
E.~Witten,
``Anti-de Sitter space and holography,''
Adv.\ Theor.\ Math.\ Phys.\  {\bf 2}, 253 (1998)
[arXiv:hep-th/9802150].
}
\lref\FreedmanTZ{
D.~Z.~Freedman, S.~D.~Mathur, A.~Matusis and L.~Rastelli,
``Correlation functions in the CFT($d$)/AdS($d+1$) correspondence,''
Nucl.\ Phys.\ B {\bf 546}, 96 (1999)
[arXiv:hep-th/9804058].
}

\lref\IW{
K.~Intriligator and B.~Wecht,
``The exact superconformal R-symmetry maximizes a,''
arXiv:hep-th/0304128.
}

\lref\HM{
C.~Herzog and J.~McKernan, ``Dibaryon Spectroscopy,'' hep-th/0305048.
}

\lref\SYZ{
A.~Strominger, S.~T.~Yau and E.~Zaslow,
``Mirror symmetry is T-duality,''
Nucl.\ Phys.\ B {\bf 479}, 243 (1996)
[arXiv:hep-th/9606040].
}

\lref\DouglasDE{
M.~R.~Douglas, B.~R.~Greene and D.~R.~Morrison,
Nucl.\ Phys.\ B {\bf 506}, 84 (1997)
[arXiv:hep-th/9704151].
}

\lref\AharonyTI{
O.~Aharony, S.~S.~Gubser, J.~M.~Maldacena, H.~Ooguri and Y.~Oz,
``Large N field theories, string theory and gravity,''
Phys.\ Rept.\  {\bf 323}, 183 (2000)
[arXiv:hep-th/9905111].
}

\lref\STD{
B.~Feng, S.~Franco, A.~Hanany and Y.~H.~He,
``Symmetries of toric duality,''
JHEP {\bf 0212}, 076 (2002)
[arXiv:hep-th/0205144].
}

\lref\BGLP{
C.~Beasley, B.~R.~Greene, C.~I.~Lazaroiu and M.~R.~Plesser,
``D3-branes on partial resolutions of abelian quotient singularities of  Calabi-Yau threefolds,''
Nucl.\ Phys.\ B {\bf 566}, 599 (2000)
[arXiv:hep-th/9907186].
}

\lref\FrancoAE{
S.~Franco and A.~Hanany,
``Geometric dualities in 4d field theories and their 5d interpretation,''
JHEP {\bf 0304}, 043 (2003)
[arXiv:hep-th/0207006].
}

\

\def\drawbox#1#2{\hrule height#2pt 
        \hbox{\vrule width#2pt height#1pt \kern#1pt \vrule width#2pt}
              \hrule height#2pt}

\def\Asym#1#2{\vcenter{\vbox{\drawbox{#1}{#2}
              \kern-#2pt       
              \drawbox{#1}{#2}}}}

\Title{\vbox{\baselineskip12pt\hbox{hep-th/0305046}
\hbox{UCSD-PTH-03-07} }} 
{\vbox{\centerline{Baryon Charges in 4d Superconformal
Field Theories}
\medskip\centerline{and Their AdS Duals}}}
\centerline{Ken Intriligator and Brian Wecht}
\bigskip
\centerline{Department of Physics} \centerline{University of
California, San Diego} \centerline{La Jolla, CA 92093-0354, USA}

\bigskip
\noindent
We consider general aspects of the realization of R and non-R flavor
symmetries in the $AdS_5\times H_5$ dual of 4d $\N =1$ superconformal
field theories.  We find a general prescription for computing the
charges under these symmetries for baryonic operators, which uses only
topological information (intersection numbers) on $H_5$.  We find and
discuss a new correspondence between the nodes of the SCFT quiver diagrams
and certain divisors in the associated geometry.  We also discuss connections
between the non-R flavor symmetries and the enhanced gauge symmetries 
in non-conformal theories obtained by adding wrapped branes.

\Date{May 2003}

\newsec{Introduction}

Interesting gauge theories arise in string theory from D-branes at
geometric singularities.  This was first studied for orbifold
singularities \refs{\DM, \DouglasDE} and, more recently, for more
general singularities.  In this paper, we will be especially
interested in 4d $\N =1$ superconformal field theories (SCFTs) which
can be string engineered by placing $N$ D3 branes at the conical singularity
of a general (local) Calabi-Yau 3-fold $X_6$.  For large $N$, it is
useful to consider the AdS dual \AharonyTI\ description of these
SCFTs, which is IIB string theory on $AdS_5\times H_5$ with $H_5$ the
horizon manifold \DMRP\ of $X_6$.  A mirror
IIA construction is to wrap $N$ D6 branes on the SYZ \SYZ\ $T^3$ of
the mirror geometry $\widetilde X_6$.

String theory and AdS/CFT provide useful insights into the SCFTs which
can be so constructed.  For example, AdS/CFT implies that the D3 brane
world-volume gauge theory actually does flow to an interacting RG
fixed point in the IR.  It is thus interesting to study generally
which quiver gauge theories can be string engineered, and what sorts
of general predictions string theory makes about these SCFTs.

Many techniques have been developed over the past several years to
help determine, given a general singularity $X_6$, precisely what is
the associated world-volume quiver gauge theory (or rather theories,
since there can be many Seiberg dual descriptions of the same
superconformal field theory).  As of yet, however, there is no
completely general method to systematically answer this question. A
partial answer can be found via partial resolution of orbifold
singularities \BGLP, but there is no systematic method for following the
RG flow from the orbifold SCFT to that of the partially resolved
singularity.  Another method, which is useful for toric singularities,
is to go to the mirror IIA description, where the gauge group and
matter content can be determined in terms of intersections of 3-cycles
\refs{\HV, \HIV, \IH}.  
Still another method, which seems to give correct results \MW\ even
outside of its expected regime of validity, is to analyze the IIB
D3-brane gauge theories (``bundles on cycles'') in the large volume
limit, and then just extrapolate to the opposite limit of singular
$X_6$.  For a selection of additional relevant references, see
\refs{\BH, \CBRP, \CFIKV, \BHK, \STD, \FrancoAE, \FFHH}. 

We will here study some general aspects of the world-volume
SCFTs which can be constructed via IIB D3-branes at singularities and
aspects of their AdS duals, focusing particularly on the geometric
realization of the flavor symmetries of SCFTs.  We will here
only consider cases where the world-volume gauge theory has already
been determined by the above mentioned methods, but we hope that some
of the methods we discuss could also be helpful in determining
the world-volume gauge theories for more general singularities. 

The world-volume gauge theories thus obtained are of the general quiver form
\eqn\gaugeg{\prod _\alpha U(Nd_\alpha),}
where $\alpha$ run over the nodes of the quiver.  The quiver
and coefficients $d_\alpha$ depend on the particular
singularity.  The theories are generally chiral, with $n_{\alpha 
\beta}>0$ chiral superfield bifundamentals $Q_{\alpha
\beta}^i$, $i=1\dots n_{\alpha \beta}$, in the $(Nd_\alpha ,
\overline{Nd_\beta})$ of $U(N d_\alpha )\times U(N d_\beta)$. 
We take $n_{\beta \alpha}=-n_{\alpha \beta}$, and draw the arrows on
the quiver so that $n_{\alpha \beta}>0$ means that the arrow(s) point
{}from node $\alpha$ to node $\beta$.  Absence of gauge anomalies
requires each node to have the same number of incoming and outgoing
arrows, so $\sum _\beta n_{\alpha \beta}d_\beta =0$ for every
$\alpha$.

The SCFT is specified by the quiver diagram and the superpotential,
which is also determined from the geometry and is a sum of terms of
the form
\eqn\wis{W=a_{i_1\dots i_k}\Tr Q^{i_1}_{\alpha _1 \alpha _2}\dots
Q^{i_l}_{\alpha _l \alpha _{l+1}}\dots Q^{i_k}_{\alpha _k \alpha _1}}
with the bifundamental gauge indices contracted in a closed loop to form
a gauge invariant meson of the quiver theory.  There are also 
gauge invariant baryons, but these generally do not enter into 
the superpotential.

We will be interested in studying the bifundamentals $Q_{\alpha
\beta}$ and their charges under the flavor symmetries.  In the AdS
dual, we only see the gauge invariant operators; in particular, we see
the baryons formed from the bifundamentals rather than the
bifundamentals themselves.  To simplify our discussion, we will here
only consider cases where all $d_\alpha =1$ in \gaugeg, so the
baryons are simply $B_{\alpha \beta}=\det _{N\times N} (Q_{\alpha
\beta})$.  Thus, the charges of the $B_{\alpha \beta}$ are just a
factor of $N$ times those of the $Q_{\alpha \beta}$.
In AdS/CFT the baryonic operators map to particles in the
$AdS_5$ bulk, which arise as D3 branes wrapping 3-cycles of $H_5$.

The 4d $\N =1$ SCFT has a global symmetry group $SU(2,2|1) \otimes
\F$, where $SU(2,2|1)$ contains the superconformal $U(1)_R$ symmetry
whose existence is necessary for a SCFT and where $\F$ are non-R
flavor symmetries.  We will be especially interested in a $U(1)^n$
subgroup of $\F$; these are the flavor symmetries under which all
$n_{\alpha \beta}$ bifundamentals $Q_{\alpha \beta}$ carry the same
charge, so the baryons are charged under these.  In the AdS dual, the
continuous global symmetries are all gauge symmetries in the $AdS_5$
bulk.  In particular, $U(1)_R$ arises as a Kaluza-Klein gauge field
coming from the metric; it is associated with a geometric isometry of
the horizon manifold $H_5$.  The $U(1)^n\subset \F$ gauge fields in
$AdS_5$ arise via reduction of the IIB RR gauge field $C_4$ on $n$
independent 3-cycles of $H_5$.  Since the baryons are wrapped D3s,
they are charged under these gauge fields.

Supposing that $H_5$ is a regular Einstein-Sasaki manifold (this
assumption might not actually be necessary for our discussion to
apply), it can be written as a $U(1)$ fibration over a four
dimensional surface $V_4$
\BG.  The $U(1)$ fiber is the isometry associated with the $U(1)_R$
symmetry, and the baryons $B_{\alpha \beta}$ must wrap this fiber
since they are charged under the superconformal $U(1)_R$ \refs{
\CBRP, \BHK}.  In addition,
the baryons wrap certain holomorphic 2-cycles $L_{\alpha \beta} \subset
V_4$.  The holomorphic condition on the 2-cycles is necessary for the
3-cycle obtained via including the $U(1)_R$ fiber to be
supersymmetric.  

Therefore the baryons, and thus also the bifundamentals $Q_{\alpha \beta}$
in our quiver gauge theory, are associated with divisors $L_{\alpha
\beta}$ on $V_4$.  All $n_{\alpha \beta}$ bifundamentals connecting nodes
$\alpha$ and $\beta$ are associated
with the same divisor $L_{\alpha \beta}$, and we take $L_{\alpha \beta}=
L_{\beta \alpha}$.  As far as we know, a general method for determining
the correct $L_{\alpha
\beta}$ has not appeared in the literature, though they were discussed
in detail for a particular example, $V_4=dP_3$, in \CBRP.  As we
discuss, the $U(1)_R$ and flavor charges of the $Q_{\alpha \beta}$ are
determined via topological intersections with the corresponding
$L_{\alpha \beta}$.  For example, the $U(1)_R$ charge of the baryons
is related to their dimension via $R[B_{\alpha \beta}] ={2\over
3}\Delta [B_{\alpha \beta}]$, which is proportional to the volume of
the 3-cycles which the baryon wraps \refs{\CBRP , \BHK}.
This yields (when all $d_\alpha =1$ in \gaugeg)
\eqn\rchargeis{R[Q_{\alpha \beta}]={2c_1\cdot L_{\alpha \beta}
\over c_1\cdot c_1},}
measured by the intersection of the
divisor with the first Chern class of $V_4$.

The $U(1)^n$ non-R flavor charges in $\F$ are given by all possible 
independent divisors
$J_i$ of $V_4$ which are orthogonal to the first Chern class of $V_4$: 
\eqn\jorth{J_i\cdot c_1=0, \qquad i=1\dots n.}
This condition, via
\rchargeis, is required for the flavor current to be $U(1)_R$ neutral.
We can pick an arbitrary basis of such $J_i$, satisfying $J_i\cdot
c_1=0$.  The charges of the bifundamentals under these flavor
symmetries is
\eqn\fchargeis{F_i[Q_{\alpha \beta}]=J_i\cdot L_{\alpha \beta}.}
While the overall normalization of the R-symmetry is fixed, that of 
the other flavor symmetries is irrelevant. 

It is interesting that string theory ``knows'' which is the correct
superconformal $U(1)_R$ symmetry, i.e. precisely which $U(1)_R$ is the
one which is in the same supermultiplet as the stress tensor.  In the
geometry, this preferred $U(1)_R$ is precisely that which is measured
by $c_1$, rather than some linear combination of $c_1$ and the $J_i$.
Finding the correct superconformal $U(1)_R$ directly via field theory
methods was, until recently, an open problem.  Inspired by our
geometric results discussed here, we have very recently found \IW\ a field
theory method to determine the superconformal $U(1)_R$.  We will verify
in examples that our field theory condition \IW\ agrees with the result
\rchargeis.   

We find some interesting properties which the divisors $L_{\alpha
\beta}$, which describe the bifundamentals in the quiver, must
satisfy.  We now summarize these results for the simplifying case
where all $d_\alpha =1$ in \gaugeg.  First, our 
$U(1)_R$ and flavor symmetries \rchargeis\ and \fchargeis\ must
not have any ABJ anomalies.  This is equivalent to the
requirement that, for every node $\alpha$, we must have
\eqn\rsymanom{\sum _{\beta} |n_{\alpha \beta}|L_{\alpha \beta}=
(N_f(\alpha)-1)c_1.}  $N_f(\alpha)$ is the total number of
flavors at node $\alpha$: $N_f (\alpha) =\half
\sum _{\beta}|n_{\alpha \beta}|$.  In addition, the superpotential
must respect these charges.  This implies that every term in the
superpotential must have net divisor equal to $c_1$, since then
\rchargeis\ and \fchargeis\ properly assign the superpotential
R-charge 2 and flavor charge 0.  Hence, for every
non-zero superpotential term, $\Pi_{\alpha \beta}Q_{\alpha
\beta}^{m_{\alpha
\beta}}$, we must have 
\eqn\superr{\sum _{\alpha \beta}m_{\alpha\beta}L_{\alpha \beta}=c_1.} 

Furthermore, we find that the $L_{\alpha \beta}$ can be written as
differences of divisors $L_\alpha$, which are associated with the
nodes of the quiver:
\eqn\labis{L_{\alpha \beta}={n_{\alpha \beta}\over |n_{\alpha
\beta}|}(L_\beta- L_\alpha) + 
c_1\theta _{\alpha \beta}\quad\hbox{where}\quad \theta _{\alpha \beta}
\equiv
\cases{0& if ${n_{\alpha \beta}\over |n_{\alpha \beta}|}
(L_\beta - L_\alpha )\geq 0$\cr 1 &if ${n_{\alpha \beta}\over |n_{\alpha
\beta}|}(L_\beta - L_\alpha) <0.$}}
By taking $\beta$ to be the endpoint of the 
$|n_{\alpha \beta}|$ arrows, and $\alpha$ the start, the factors $n_{\alpha
\beta}/|n_{\alpha \beta}|$ become $+1$.   The sign of $L$ in \labis\ refers
to the sign of $c_1\cdot L$, and we'll always choose the $L_\alpha >0$
in this sense.  We require that all $L_{\alpha \beta}\geq 0$ because
the expression \rchargeis\ must assign non-negative R-charge to all
chiral superfields.  For most $L_{\alpha \beta}$, the $\theta _{\alpha
\beta}$ term in
\labis\ vanishes.  In fact, every term in the superpotential \wis\ has
precisely one $Q_{\alpha \beta}$ for which the associated $\theta _{\alpha 
\beta} =1$, with the others having $\theta _{\alpha \beta}=0$, and this
ensures that \superr\ is satisfied: 
every superpotential term has net divisor $c_1$.

The anomaly free condition \rsymanom\ implies that, for every node
$\alpha$ of the quiver,
\eqn\anomx{\sum _\beta n_{\alpha \beta}L_\beta = \sum _{outgoing\
\beta}|n_{\alpha \beta}|L_\beta - \sum _{incoming\ \beta}|n_{\alpha
\beta}|L_\beta =0 \quad\hbox{mod}
\quad c_1;}
we could write the specific coefficient of $c_1$ on the RHS in terms of
$N_f(\alpha)$ and the $\theta _{\alpha \beta}$, but \anomx\ suffices for
a later application. Outgoing $\beta$ means those nodes where the arrow
goes out from $\alpha$, toward $\beta$.  We used the fact that, mod $c_1$,
$0=\sum _\beta |n_{\alpha \beta}|L_{\alpha \beta}=\sum _\beta n_{\alpha \beta}
(L_\beta - L_\alpha)$, and $\sum _\beta n_{\alpha \beta}=0$.  

Using \labis, the superconformal $U(1)_R$ charges \rchargeis\ and 
other flavor charges \fchargeis\ can be expressed as differences
of charges associated with the nodes of the quiver:
\eqn\rchargeiss{\eqalign{R[Q_{\alpha \beta}]&=R(\beta) - R(\alpha) +
2\theta _{\alpha \beta}, \qquad R(\alpha ) \equiv {2c_1\cdot L_\alpha
\over c_1\cdot c_1},\cr F_i[Q_{\alpha \beta}]&=F_i(\beta )-F_i(\alpha)
\qquad \qquad \qquad F_i(\alpha)
\equiv J_i\cdot L_\alpha.}}

 As we discuss in section 2, the $L_\alpha$ in \labis\ are expected to
have some natural mathematical meaning, in some way related to a {\it
dual} version of the collection of bundles on divisors $\O
(D_\alpha)$.  However, we were not able to make this precise here, and
did not find a fully general method to independently obtain the
$L_\alpha$ from first principles.

As we also discuss, Seiberg duality \SeibergPQ\ has a simple action on the 
$L_\alpha$.  To simplify the discussion, we consider the case where the
dualized gauge group at node $\alpha$ has $N_f=2N_c$, so
that the rank of the dualized gauge group is the same as it was originally;
then all $d_\alpha =1$, for both the original and also the dualized quivers.  
We write the bifundamentals associated with node $\alpha$ as
$Q_{\alpha \beta}$, $Q_{\alpha \gamma}$, $Q_{\rho \alpha}$, and 
$Q_{\sigma \alpha}$ with the arrows going out from node $\alpha$ out to 
$\beta$ and $\gamma$ (which could be the same node) and into node
$\alpha$ from $\rho$ and $\sigma$ (which could also be the same).  
The dualized quiver has dual quark bifundamentals, with reversed
arrows, and also bifundamentals corresponding to the mesons of the
original $U(N)_\alpha$ theory.  We show that the duality correspondences
and R-charge and flavor charge assignments imply that Seiberg dualizing
node $\alpha$ only changes the $L_\alpha$ of that node, as 
\eqn\sduall{L_{\alpha}^{\prime} = L_\beta + L_\gamma - L_\alpha,}
with the $L$'s associated with the other nodes 
remaining unchanged after Seiberg duality.     

One can also construct non-conformal theories, e.g. by wrapping D5
branes on cycles $\Sigma _i \subset X_6$, with the other directions
filling the uncompactified 4d space transverse to $X_6$.  As discussed
in \CFIKV, there is a flux condition which requires that the
two-cycles $\Sigma _i$ of $X_6$ not intersect any compact 4-cycles
(this condition also rules out wrapping D7s on 4-cycles).  The cycles
$\Sigma _i$ which the D5's wrap correspond to divisors in $V_4$, and
the flux condition implies that they must have zero intersection with
$c_1(V_4)$.  Thus the total 5-brane charge must be that of $N_i$ D5s
wrapped on divisors $J_i$ of $V_4$, i.e. $\sum _i N_i J_i$ where every
$J_i$ satisfies $c_1\cdot J_i=0$.  These are the same $J_i$ in \jorth,
corresponding to the non-R flavor symmetries of the SCFT theory
without wrapped D5s.  Indeed, the flavor symmetries of the SCFT
without wrapped D5s become part of the gauge symmetry in the theory
with wrapped D5's:
\eqn\gaugegw{\prod _{\alpha} U(N+M_\alpha) \qquad \hbox{with}
\qquad M_\alpha = \sum _i N_i J_i \cdot L_\alpha.}

Because the flavor charges \fchargeis\ of the bifundamentals have
become part of their gauge charge in the theory \gaugegw\ with added
wrapped D5s, consistency of the theory \gaugegw\ requires that the
flavor symmetries $F_i$ have vanishing 't Hooft anomalies
\eqn\vanihooft{\Tr F_i =0 \qquad \hbox{and}\qquad
\Tr F_iF_j F_k =0 \qquad \hbox{for all}
\qquad i,j,k.}
This can be seen to be the case from the origin of these symmetries in
the $AdS_5\times H_5$ dual, as the reduction of $C_4$ on 3-cycles of
$H_5$: the $C_4$ gauge field does have the particular Chern-Simons
type terms which would be needed to yield non-zero 't Hooft anomalies
upon reduction on $H_5$.  

The outline of this paper is as follows: In section 2, we discuss how
our main results, reviewed above, are obtained.  In section 3, we
discuss aspects of 't Hooft anomalies and our field theory result \IW\ 
for determining the superconformal $U(1)_R$.  In section 4, we
illustrate our ideas for the examples of certain toric and non-toric
del Pezzo surfaces.  We expect that the methods apply more generally.

While this paper was in preparation, Chris Herzog and James McKernan
alerted us to their related work \HM.  

{\bf Note added in revised version, July '03:} Several of the loose
ends raised in this paper were subsequently analyzed and clarified in
a nice paper by Herzog and Walcher \ref\HW{C.P. Herzog and J. Walcher,
``Dibaryons from Exceptional Collections,'' hep-th/0306298}.  Among
other things, they presented a precise notion of the ``dual'' to the
exceptional collection, which is related to the $L_\alpha$ that we
introduced in
\labis. 

\newsec{Some string predictions}

One way to find the quiver gauge theory associated with a singularity is
in terms of a collection of sheaves.  These often can be
written as  $\O (D_\alpha)$, where $D_\alpha$ is some set of divisors of
$V_4$.  Given such an collection,
the number of bifundamentals can then be computed by
the formula 
\eqn\bifund{n_{\alpha \beta} = \chi ( \O (D_\alpha),\O(D_\beta)) -
\chi (\O(D_\beta),\O(D_\alpha)),} 
where 
$\chi ( \O (D_\alpha),\O(D_\beta))=
\sum_{i=0}^3(-1)^i{\rm dim}{\rm Ext}^i(\O (D_\alpha),\O(D_\beta))$ 
is the relative Euler characteristic of the two sheaves. (For an
{\it exceptional} collection one of the two terms in \bifund\ vanishes.)  A
small modification of \bifund, though, is generally needed: at certain
nodes, the directions of the bifundamentals need to be flipped.  This
is seen in the gauge theory because otherwise some gauge groups would
be anomalous.  The flip is a continuation of $N\rightarrow -N$ for the
corresponding node.  Precisely which nodes require such a flip can be
determined by the methods of \refs{\HV, \HIV}.

We consider the situation where all gauge groups at nodes $\alpha$ are
$U(N)$, to simplify the analysis of the 
baryons $B_{\alpha \beta}=\det _{N\times N}(Q_{\alpha
\beta})$.  If there is a multiplicity $n_{\alpha \beta}>1$ of bifundamentals,
there will be a corresponding multiplicity of baryons; we do not
consider such baryon multiplicities further.  In AdS/CFT, the baryons
arise as D3 branes wrapped on 3-cycles of $H_5$, with the dimension of
the corresponding operator directly proportional to the volume of the
corresponding 3-cycle, see \refs{\CBRP, \BHK}.  This yields
$R(B_{\alpha \beta})= {2\over 3}\Delta (B_{\alpha
\beta})={2\over 3}\mu _3 L^4\V (\Sigma ^3_{\alpha \beta})$, where $\mu_3$
is the tension of the brane.  The $\Sigma ^3_{\alpha \beta}$ 
corresponds to a holomorphic divisor $L_{\alpha \beta}$ of $V$,
combined with the $S^1$ fiber.  As discussed in 
\refs{\CBRP, \BHK}, we have 
\eqn\voli{\V (H_5)=\left({2\pi q\over 3}\right) \V (V)=
\left({\pi ^3 q\over 27}\right)c_1\cdot c_1,}
where $2\pi q/3$ is the length of the $U(1)$ fiber and $c_1$ is the
first Chern class of $V$, with $c_1\cdot c_1\equiv \int _V c_1\wedge
c_1$.  Here $q$ is defined by $c_1(V)=qc_1(U(1))$, with $c_1(V)$ the
first Chern class of the 2 complex dimensional K\"{a}hler-Einstein
manifold $V$, which satisfies $\omega = {\pi \over 3} c_1(V)$ with
$\omega$ the K\"{a}hler form of $V$, and $c_1(U(1))$ is the first
Chern class of the $U(1)$ line bundle.  Then $R(B_{\alpha
\beta})={2\over 3}\mu _3L^4 \V (\Sigma ^3_{\alpha \beta})$ yields
\eqn\rbaryis{R(B_{\alpha \beta})=\left({2\over 3}\right)
\left({N\pi \over 2\V(H_5)}\right)\left({2\pi q \over 3}\right)
{\pi \over 3} c_1 \cdot L_{\alpha \beta} =
2N{c_1\cdot L_{\alpha \beta}\over c_1\cdot c_1}.}
Since $R[B_{\alpha \beta}]=NR[Q_{\alpha \beta}]$, this yields \rchargeis.  

The non-R flavor symmetries $\F$ under which the baryons are charged
come from reducing the IIB gauge field $C_4$ on 3-cycles of $H_5$.
These 3-cycles must include the $U(1)$ fiber direction, along with
some divisors $J_i$ of $V_4$.  Since these flavor symmetries must be
R-neutral, \rchargeis\ implies that the $J_i$ must satisfy $c_i\cdot
J_i=0$.  A similar consideration as in \rbaryis\ then leads to the 
flavor charge assignments of the baryons and hence the bifundamentals,
as in \fchargeis.  Again, the overall normalization of these non-R
$U(1)$ flavor charges is irrelevant, so we drop the normalization factor
of $2/(c_1\cdot c_1)$ for these.  

The condition that the $U(1)_R$ symmetry be anomaly free is
that, at every node $\alpha$,  
\eqn\anomf{2d_\alpha +\sum _\beta |n_{\alpha \beta}|(R[Q_{\alpha \beta}]-1)
d_\beta =0.}
In our situation, where all $d_\alpha =1$, this together 
with \rchargeis\ requires
\eqn\ii{c_1\cdot \sum _\beta |n_{\alpha \beta}|L_{\alpha
\beta}=(N_f(\alpha)-1)c_1\cdot c_1,}
where $N_f(\alpha)=\half \sum _\beta |n_{\alpha \beta}|$ is the
number of flavors at node $\alpha$. Likewise, the condition that the
$U(1)_i$ flavor symmetries associated with $J_i$ have vanishing ABJ
anomaly at every node $\alpha$ is
\eqn\anomff{\sum _\beta |n_{\alpha \beta}|F_i[Q_{\alpha \beta}]=0.}
This, together with \fchargeis, implies that
\eqn\iii{J_i\cdot \sum _\beta |n_{\alpha \beta}| L_{\alpha \beta}=0
\qquad\hbox{for all}\qquad J_i\cdot c_1=0.}
Taken together, \ii\ and \iii\ imply \rsymanom, which is a very
restrictive condition on the $L_{\alpha \beta}$.

In addition, the $L_{\alpha \beta}$ must satisfy another condition in
order that the superpotential respect the $U(1)_R$ and flavor
symmetries.  Since every term in the superpotential must have R-charge
2 and non-R flavor charge zero, the total divisor associated with any
superpotential term must be precisely $c_1$. Thus a necessary 
(but generally not sufficient)
condition for a non-zero superpotential term $\Pi_{\alpha
\beta}Q_{\alpha \beta}^{m_{\alpha \beta}}$ is
\eqn\superr{\sum _{\alpha \beta}m_{\alpha\beta}L_{\alpha \beta}=c_1.} 

We have found that, furthermore, the $L_{\alpha \beta}$ can be written
as a difference of divisors $L_\alpha$ which are associated with the
nodes of the quiver, as in \labis.  The condition \labis\ is sufficiently 
restrictive so that, given the
$L_{\alpha \beta}$, the $L_\alpha$ can be determined -- up to
the addition of an overall constant divisor to all $L_\alpha$, which
would cancel on the RHS of \labis.    

We expect that the $L_\alpha$
must have a natural mathematical interpretation, which could be used
to independently determine them.  To get some insight into 
what this direct interpretation of the $L_\alpha$ might be, 
consider the process of partially
resolving the geometric singularity.  This  corresponds to turning
on a FI term at some node, which forces some bifundamentals to then
have a non-zero expectation value, Higgsing the world-volume
gauge theory down to that of the resolved singularity.  Thus the
bifundamental divisors $L_{\alpha\beta}$ are naturally associated
with differences of FI terms at the nodes $\alpha$ and $\beta$:
roughly, $L_{\alpha \beta}\sim \zeta _\beta - \zeta _\alpha$.  
We interpret this as in \labis. Note
that the FI terms are dual to the bundles at the nodes,
since we have a corresponding coupling $\int d^4 x d^4 \theta
\zeta _\alpha V_\alpha$, so this suggests that the $L_\alpha$ are
dual\foot{We thank M. Douglas for suggesting this to us.} 
to the $\O (D _\alpha)$.  We do not presently, however, have a
prescription for how to make this precise.  

So, for the present work, we found the $L_\alpha$ on a case-by-case
basis by first obtaining the $L_{\alpha\beta}$, and then using
\labis.  We have found that, at least in all those cases which we have
considered, it is possible to take $L_\alpha = D_\alpha$ for most, but
not all, of the nodes.  In particular, in the examples that we
considered, we can take $L_\alpha = D_\alpha$ for all of those nodes
for which no $N\rightarrow -N$ flip is required to get the
bifundamentals via
\bifund.  For those nodes which do require such a flip, the $L_\alpha
\neq D_\alpha$ and, since we do not yet know the general
procedure for determining these $L_\alpha$, we determined them on a
case-by-case basis in the examples by imposing the very restrictive
consistency conditions, discussed above, which the $L_{\alpha \beta}$
must satisfy.

As mentioned in the introduction, we can also consider non-conformal
theories, obtained by wrapping $N_i$ D5 branes on 2-cycles of $X_6$.  Doing so
leads to a quiver of the same form as in the conformal case, but with
the gauge group modified, as in \gaugegw; as indicated, the $L_\alpha$
determine the gauge group modification:
\eqn\gaugegww{\prod _\alpha U(N+M_\alpha)\qquad\hbox{with}
\qquad M_\alpha = \sum _i N_i J_i \cdot L_\alpha.}
Absence of gauge anomalies at every node $\alpha$ requires
\eqn\nogaugea{\sum _\beta n_{\alpha \beta}(N+M_\beta)=0,}
which is indeed satisfied thanks to \anomx, since $J_i\cdot c_1=0$.  

These theories with the wrapped D5s effectively gauge our previously
global $U(1)^n$ flavor symmetries.  
Consider, as an example, the case where $N_1=1$ and all other $N_i=0$.
The gauge group is then $\prod _\alpha U(N+J_1\cdot L_\alpha)$,
which has as a subgroup $U(1)\times \prod _\alpha U(N)$, with 
bifundamentals $Q_{\alpha \beta}$ having charge $J_1\cdot (L_\beta
-L_\alpha)$ under the $U(1)$.  The additional gauged $U(1)$ here
is just the flavor $U(1)$ associated with $J_1$.  Likewise, the 
general gauge theory \gaugegww\ has as a subgroup $U(1)^n\times \prod
_\alpha U(N)_\alpha$, where the $U(1)^n$ correspond to what were flavor
symmetries before we added the wrapped D5s.  

There is another condition on the $M_\alpha$ appearing above, which is
required by the flux condition discussed e.g. in \CFIKV:
\eqn\msumi{c_1\cdot \sum _\alpha (N+M_\alpha)\epsilon _\alpha D_\alpha =0,}
where $\epsilon _\alpha = +1$ for all nodes except for those which 
require a $N\rightarrow -N$ sign flip in \bifund; for these flipped nodes,
$\epsilon _\alpha =-1$.  Considering the case of no wrapped D5s, the
$D_\alpha$ must satisfy 
\eqn\dsumi{c_1\cdot \sum _\alpha \epsilon _\alpha
D_\alpha =0 \qquad\hbox{thus}\qquad
\sum _\alpha \epsilon _\alpha
D_\alpha = \sum _i \widetilde N_i J_i.}
If we took for the $\epsilon _\alpha D_\alpha$ the ``fractional
brane'' charges of the sort discussed e.g. in \IH, we would have 
$\widetilde N_i=0$ in \dsumi, but generally we will not
make that choice, instead taking the $D_\alpha$ to satisfy the weaker
condition \dsumi.  Including wrapped D5s, we must have $\sum _\alpha 
M_\alpha \epsilon _\alpha c_1\cdot D_\alpha =0$, implying that $\sum
_\alpha \epsilon _\alpha (J_i\cdot L_\alpha)D_\alpha$ is 
in the span of the $J_i$.  

Finally, consider the action of Seiberg duality \SeibergPQ\ on a node
$\alpha$, which we suppose has $N_f=2N_c$ in order for the gauge group
to be self-dual. We also suppose that all $d_\alpha =1$ in \gaugeg.
We write the bifundamentals associated with node $\alpha$ as
$Q_{\alpha \beta}$, $Q_{\alpha \gamma}$, $Q_{\rho \alpha}$, and 
$Q_{\sigma \alpha}$ with the arrows going out from node $\alpha$ out to 
$\beta$ and $\gamma$ (which could be the same node) and into node
$\alpha$ from $\rho$ and $\sigma$ (which could also be the same).  
The dualized quiver has dual quark bifundamentals, with reversed
arrows, and also bifundamentals corresponding to the mesons of the
original $U(N)_\alpha$ theory.  We write the dual quarks as 
$Q'_{\beta \alpha}$, $Q'_{\gamma \alpha}$, $Q'_{\alpha
\rho}$, and $Q'_{\alpha \sigma}$.  The bifundamentals coming from
the mesons are $Q'_{\rho \beta}=Q_{\rho \alpha}Q_{\alpha \beta}$,
$Q'_{\sigma \beta}= Q_{\sigma \alpha}Q_{\alpha \beta}$, $Q'_{\rho
\gamma}=Q_{\rho \alpha}Q_{\alpha
\gamma}$, and $Q'_{\sigma \gamma}=Q_{\sigma \alpha}Q_{\alpha \gamma}$.
Since the R-charge and flavor charges, given by \rchargeis\ and \fchargeis,
must respect this map, the divisors associated with the meson legs of the dual 
quiver must satisfy
\eqn\dualmi{\eqalign{L'_{\rho \beta}&=L_{\rho \alpha}+L_{\alpha \beta}, 
\qquad L'_{\sigma \beta}=L_{\sigma \alpha}+L_{\alpha \beta}, \cr
L'_{\rho \gamma}&=L_{\rho \alpha}+L_{\alpha \gamma}, \qquad L'_{\sigma
\gamma}=L_{\sigma\alpha}+L_{\alpha \gamma}.}}

Duality maps the baryons of the original theory to those of the
dual, as $Q_{f_1}\dots
Q_{f_{N_c}}= \epsilon _{f_1\dots f_{N_f}}q^{f_{N_c+1}}\dots q^{f_{N_f}}$
\SeibergPQ.  For our theory, this implies a mapping $\det Q_{\alpha
\beta}= \det Q'_{\gamma \alpha}$, and hence the map $Q_{\alpha \beta}
\leftrightarrow Q'_{\gamma \alpha}$.  We reversed the direction
of the arrows, because the dual quarks transform in the conjugate flavor
representation (and we then need to apply charge conjugation on node
$\alpha$ to get back bifundamentals).  We exchanged the $\beta$ and $\gamma$
because of the $\epsilon _{f_1\dots f_{N_f}}$ in the baryon map, which
maps e.g. $\det Q_{\alpha \beta}$ to $\det q_{\gamma \alpha}$.  Likewise,
the other bifundamentals map as $Q_{\alpha \gamma}
\leftrightarrow Q'_{\beta \alpha}$, $Q_{\rho \alpha}\leftrightarrow
Q'_{\alpha \sigma}$, and $Q_{\sigma \alpha}\leftrightarrow Q'_{\alpha
\rho}$.  Since the R-charge and flavor charge assignments, given by 
\rchargeis\ and \fchargeis, must respect this map, the divisors
associated with the dual quark legs of the dual quiver must satisfy
\eqn\dualqi{L'_{\gamma \alpha}=L_{\alpha \beta}, \quad L'_{\beta
\alpha}=L_{\alpha \gamma}, \quad L'_{\alpha \rho}=L_{\sigma \alpha},
\quad L'_{\alpha \sigma}=L_{\rho \alpha}.}

The dual theory has superpotential terms such as $W={1\over 
\mu }Q'_{\rho
\beta}Q'_{\beta \alpha}Q'_{\alpha \rho}+\dots$, which must respect the 
$U(1)_R$ and flavor symmetries, and hence must have total divisor
$c_1$:
\eqn\wdualr{L'_{\rho \beta}+L'_{\beta \alpha}+L'_{\alpha
\rho}=c_1.}

Finally, all of the other nodes and legs of the original quiver are 
otherwise untouched by the Seiberg duality on node $\alpha$, so their 
charges, and hence leg divisor assignments, are the same in the dual
as in the original theory.  

All of these conditions can be satisfied very simply in terms of our
relation \labis\ for writing the divisors of the quiver's legs in terms of
divisors associated with the nodes.  Seiberg duality only acts on the
$L_\alpha$ of the dualized node $\alpha$, with the $L$'s of all other
nodes unchanged.  The conditions \dualmi\ are then almost immediately
satisfied, though there is are apparently non-trivial conditions coming
{}from the terms proportional to $c_1$: $\theta _{\rho \beta}=\theta
_{\rho \alpha} +\theta _{\alpha \beta}$ etc.; we verified that these
conditions are indeed satisfied in all of our examples.  
The conditions in \dualqi\ are also satisfied
by $L'_\alpha$ as in \sduall, with the other node $L's$ untouched.  
For example, \labis\ gives for the first relation in \dualqi:
$L'_\alpha - L_\gamma+c_1\theta (L'_\alpha - L_\gamma)= L_\beta - L_\alpha
+c_1\theta (L_\beta - L_\alpha)$, which is indeed satisfied when
$L'_\alpha = L_\beta +L_\gamma - L_\alpha$.  Note also that using 
\dualmi, \dualqi, and $N_f(\alpha)=2$,
\wdualr\ is equivalent to \rsymanom. 

\newsec{'t Hooft anomalies}

It is useful to consider the 't Hooft anomalies of the global 
flavor symmetries $U(1)_R$ and $U(1)_{F_i}$ of the SCFTs. The
conditions \anomf\ and \anomff,  that $U(1)_R$ and $U(1)_{F_i}$ 
have vanishing ABJ anomalies at each node, ends up implying
that they also have vanishing linear 't Hooft anomalies (relevant
for coupling to gravity):
\eqn\nolth{\Tr R = \Tr F_i =0.}
This is a consequence of the quiver gauge group form \gaugeg, with
only purely chiral, bifundamental matter.  

The various cubic 't Hooft anomalies (again, taking all $d_\alpha =1$
in \gaugeg\ to simplify) are
\eqn\thooftis{\eqalign{
\Tr R^3 &= N^2\sum _\alpha 
\left(1+\half 
\sum _\beta |n_{\alpha \beta}|(R[Q_{\alpha\beta}]-1)^3\right),\cr
\Tr R F_i F_j&= \half N^2 \sum _{\alpha \beta}|n_{\alpha
\beta}|(R[Q_{\alpha\beta}]-1)F_i[Q_{\alpha \beta}]F_j[Q_{\alpha 
\beta}],\cr
\Tr R^2 F_i &=\half N^2 \sum _{\alpha \beta}|n_{\alpha
\beta}|(R[Q_{\alpha\beta}]-1)^2F_i[Q_{\alpha \beta}],\cr
\Tr F_i F_j F_k&=\half N^2 \sum _{\alpha \beta}|n_{\alpha
\beta}|F_i[Q_{\alpha \beta}]F_j[Q_{\alpha 
\beta}]F_k[Q_{\alpha \beta}].}}
We can evaluate these in terms of the geometry via \rchargeis\ 
and \fchargeis.

Interestingly, for each of the 't Hooft anomalies \thooftis, we can
also make an independent prediction.  For example, using the AdS/CFT 
prediction for the central charges $a$ and $c$ \SG\ and their relation
with the $\Tr R^3$ 't Hooft anomaly \refs{\AFGJ, \AEFJ} leads to
the prediction
\eqn\thooft{\Tr R^3={8\over 9}N^2{\V (S^5)\over
\V (H_5)},}
which we can write in terms of $q$ and $c_1\cdot c_1$ using \voli.   
(And also $\Tr R=0$, which we've already seen to indeed be the case.)
So both \thooftis\ and \thooft\ compute $\Tr R^3$ via geometric
data; hence, some mathematical identity must ensure that the two, 
apparently different, geometric computations always agree.  We do
not yet have a general understanding of this expected identity, but
check that the computations indeed agree in all of our examples.

As discussed in \IW, the superconformal $U(1)_R$ charge has the
property that, among all possibilities, it maximizes $3\Tr R^3-
\Tr R$.  If we write the most general $U(1)_R$ symmetry as
$R=R_0+\sum _i s_i F_i$, where $R_0$ is an arbitrary initial R-symmetry
and $s_i$ are real parameters, maximizing $3\Tr R^3-\Tr R$ with 
respect to the $s_i$ yields: $9\Tr R^2F_i = \Tr F_i$ and $\Tr
RF_iF_j<0$ \IW.  In the present context, where all $U(1)_{F_i}$ have
$\Tr F_i=0$, we thus must have  
\eqn\amax{\Tr R^2F_i= 0  \qquad \hbox{and}
\qquad \Tr R F_iF_j < 0,}
specifically the latter matrix in $i$ and $j$ must have all negative
eigenvalues.  We check in all cases that \thooftis, using \rchargeis\
and \fchargeis, indeed satisfies \amax.  Again, we suspect that some
mathematical identity ensures that this is indeed always the case, e.g.
the first identity in \amax\ requires 
\eqn\amaxii{\sum _{\alpha \beta}|n_{\alpha \beta}|
(2{c_1\cdot L_{\alpha \beta}\over c_1\cdot c_1}-1)^2(J_i\cdot L_{\alpha
\beta})=0,}
for all $J_i$ satisfying $c_1\cdot J_i=0$.

Finally, the flavor symmetries $F_i$ are expected to have all vanishing
cubic 't Hooft anomalies
\eqn\fic{\Tr~F_i F_j F_k =0 \qquad \hbox{for all}\qquad i,j,k.}
Any non-zero such cubic 't Hooft anomalies would require the presence
of Chern-Simons 5-forms $\sim A\wedge F\wedge F$ in the $AdS_5$ bulk
\refs{\WittenQJ, \FreedmanTZ}, but such a term does not have a
candidate 10d origin, in terms of the 10d gauge fields $C_4\sim
A\wedge \eta$ and $F_5=F\wedge \eta$, with $\eta$ a 3-form on $H_5$.  
Further, as we mentioned above, the $F_i$ flavor symmetries become
part of the gauge symmetry upon including wrapped D5s.  Hence absence
of gauge anomalies of those theories requires \fic.  Again we can check
in all examples that, indeed, 
\eqn\ficc{\sum _{\alpha \beta}|n_{\alpha \beta}|(J_i\cdot L_{\alpha
\beta})(J_i\cdot L_{\alpha \beta})(J_k \cdot L_{\alpha \beta})=0,}
for all $i,j,k$, with $J_{i,j,k}\cdot c_1=0$.

\newsec{del Pezzo examples}

Consider the case where $X_6$ is a local Calabi-Yau which is a complex
cone over the del Pezzo surface $dP_n$.  Recall that $dP_n$ is a copy of ${\bf
P}^2$ blown up at $n$ points, where $0 \leq n \leq 8$. Each blown-up
point corresponds to an exceptional divisor $E_i$, and there is also a
divisor $D$ which is the pullback of a hyperplane on ${\bf P}^2$. The 
intersection numbers of these divisors are 
\eqn\inter{D\cdot D = 1, \qquad E_i \cdot E_j = 
-\delta_{ij}, \qquad D \cdot E_i = 0,}
and the first Chern class (anti-canonical class) is 
\eqn\ciis{c_1=3D -\sum_{i=1}^n E_i.}
There are $n$ linearly independent divisors $J_i$ satisfying $J_i\cdot c_1=0$,
so there will be a non-R $U(1)^n$ flavor symmetry under which the baryons
are charged.  These $J_i$ correspond to the root lattice of the exceptional
group $E_n$, with $E_1=A_1$, $E_2=A_1+A_1$, $E_3=A_3$, $E_4=A_4$,  $E_5=D_5$,
and $E_{6,7,8}$ as expected. In particular, if we take for our basis
\eqn\jsare{J_i = E_i-E_{i+1}\quad\hbox{ for $i=1\dots n -1$,}\quad
\hbox{and} \quad J_n=D-E_1-E_2-E_3}
their intersections $J_i\cdot J_j$ are given by the $E_n$ Cartan matrix.  
The $dP_n$ automorphisms correspond to the $E_n$ Weyl reflections on
the $J_i$.

Rewriting \rbaryis\ in this language, the R-charge of a baryon
$B_{\alpha \beta}$ corresponding to a holomorphic 2-cycle $\L$ is
\eqn\barydp{R(B_{\alpha \beta}) = 2N {{c_1 \cdot \L} \over {c_1 \cdot c_1}} = 
2N {{c_1 \cdot \L} \over {9-n}}.} Since the numerator is an integer,
this implies that the R-charge of any baryon in the $dP_n$ theory is
an integer multiple of ${2N \over 9-n}$.  Also, using \thooft\ and
\voli, we get that
the cubic 't Hooft anomaly must be
\eqn\delpth{\Tr R^3={24N^2\over 9-n}.}

In the following sections, we will work out our prescription in detail
for the case of the toric del Pezzos $dP_{n\leq 3}$ and 
the non-toric del Pezzo $dP_4$. The $dP_3$ case was studied extensively
in \CBRP, so we start with that case first.  

\subsec{Cone over $dP_3$.}

The are four known field theories that arise from the cone over $dP_3$ which 
are related to each other via Seiberg duality. One of these (usually called
Model III) is described by the $U(N)^6$ theory given by the quiver in Figure 1.

\bigskip
\centerline{\epsfxsize=0.40\hsize\epsfbox{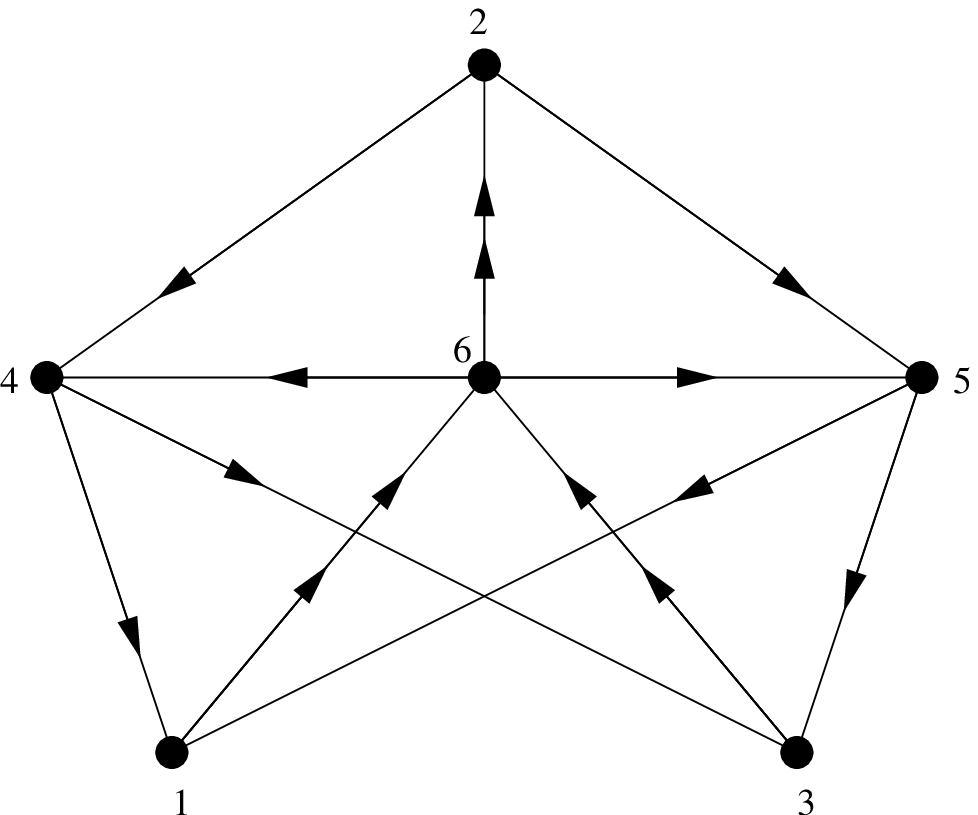}}
\centerline{\ninepoint\sl \baselineskip=8pt {\bf Figure 1:}
{\sl The Model III $dP_3$ quiver.}}
\bigskip

The correspondence between divisors and bifundamentals has already
been worked out in \CBRP\ and is given in the following table: 

\eqn\dpIII{ \matrix{{Q_{\alpha \beta}}& {L_{\alpha \beta}}\cr
X_{51} & E_1 \cr X_{24} & E_2 \cr X_{53} & E_3 \cr X_{43} & D - E_1 -
E_2 \cr X_{25} & D - E_1 - E_3 \cr X_{41} & D - E_2 - E_3 \cr X_{16} &
D - E_1 \cr X_{62} & D - E_2 \cr X_{36} & D - E_3 \cr X_{64} & D \cr
X_{65} & 2D - E_1 - E_2 - E_3 \cr}.}  Note that, because the
assignment of divisors and charges is the same for any of the
$|n_{\alpha \beta}|$ bifundamentals connecting the same nodes, we have
not explicitly written the fields $Y_{16}$, $Y_{36}$, and $Y_{62}$ in
this table. It is of course important to include these multiplicities
in all computations, e.g. when computing traces.

It is easily verified that the $L_{\alpha\beta}$ \dpIII\ indeed satisfy
our vanishing ABJ anomaly condition \rsymanom.  Furthermore, the
superpotential (found in \refs{\CBRP,\STD}) respects the symmetries,
because every term in the superpotential has exactly one field for
which $\theta _{\alpha \beta} = 1$.

We now write these $L_{\alpha \beta}$ as in \labis.  As seen in the table
below, we can take the $L_\alpha$ to equal the $D_\alpha$ which define
the collection of bundles, except at nodes 2,4,5.
These are precisely the nodes where a flip is required \CFIKV\ to 
obtain the quiver diagram; this seems to be a general connection. We
also include in the table below the $M_\alpha =\sum _i N_iJ_i \cdot
L_\alpha$, which give the modification of the gauge groups in the quiver
diagram with added wrapped D5's, as in \gaugegw.  The $J_i$ are 
as in \jsare : $J_1=E_1-E_2$, $J_2=E_2-E_3$, $J_3=D-E_1-E_2-E_3$.

\eqn\ldpii{\matrix{ {\rm Node} & L_\alpha & D_\alpha & M_\alpha \cr
1 & E_1 & E_1&N_3-N_1\cr
2 & 2D-E_2 & E_2 &N_3-N_1+N_2\cr
3 & E_3 & E_3 &N_2+N_3\cr
4 & 2D & D - E_2&2N_3\cr
5 & 0 & D - E_ 3 &0\cr
6 & D & D&N_3.}}

\noindent One can readily check that
these $L_\alpha$ and \labis\ reproduce the required $L_{\alpha \beta}$
in \dpIII.

  Since $dP_3$ has $c_1 = 3D - E_1
- E_2 - E_3$, we write the three flavor currents as $J_1 = E_2 - E_1$,
$J_2 = E_2 - E_3$, and $J_3 = D - E_1 - E_2 - E_3$. Then, we can use
\rchargeis\ and \fchargeis\ to read off the charges:

\eqn\dpIIIch{ \matrix{{Q_{\alpha \beta}} & J_1 & J_2 & J_3 & R \cr
X_{51} & -1 & 0 & 1 & 1/3 \cr
X_{24} & 1 & -1 & 1 & 1/3 \cr
X_{53} & 0 & 1 & 1 & 1/3 \cr
X_{43} & 0 & 1 & -1 & 1/3 \cr
X_{25} & 1 & -1 & -1 & 1/3 \cr
X_{41} & -1 & 0 & -1 & 1/3 \cr
X_{16} & 1 & 0 & 0 & 2/3 \cr
X_{62} & -1 & 1 & 0 & 2/3 \cr
X_{36} & 0 & -1 & 0 & 2/3 \cr
X_{64} & 0 & 0 & 1 & 1 \cr
X_{65} & 0 & 0 & -1 & 1 \cr}.}

\noindent These are exactly the $-U(1)_C$, $U(1)_D$, $-U(1)_E$, and R
charges found in \CBRP. 

Let's now examine a Seiberg dual theory, known as Model IV. The quiver
for this theory is obtained by Seiberg dualizing node 2; see Figure 2.

\bigskip
\centerline{\epsfxsize=0.40\hsize\epsfbox{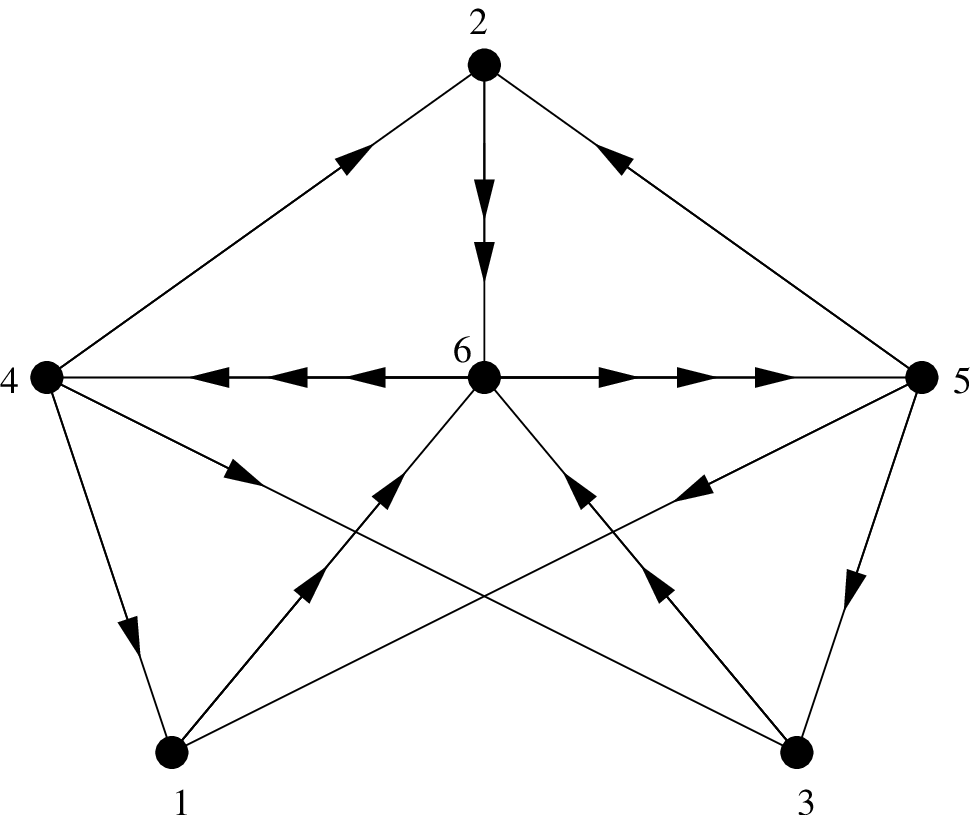}}
\centerline{\ninepoint\sl \baselineskip=8pt {\bf Figure 2:}
{\sl The Model IV $dP_3$ quiver.}}
\bigskip

The bifundamental/divisor correspondence has been worked out already in
\CBRP, so we only need to check that our prescription for 
Seiberg dualizing the $L_\alpha$ agrees. The only difference is in
$L_2$, which becomes $L_2^{\prime} = L_4 + L_6 - L_2 = E_2$, which
checks with the results of \CBRP.

\subsec{Cone over $dP_2$}

Since blowing down a divisor is equivalent to Higgsing an appropriate
bifundamental, one can easily obtain the $dP_2$ quiver by Higgsing any
bifundamental field in the $dP_3$ quiver that corresponds to an
exceptional divisor. Depending on which divisor gets Higgsed, the
resulting quiver will be one of two possible Seiberg dual theories. We
choose to blow down $E_2$ or, equivalently, Higgs $X_{24}$. The
resulting quiver appears in Figure 3 (for simplicity, we have
relabeled nodes $6 \rightarrow 4$ and $4/2 \rightarrow 2$).

\bigskip
\centerline{\epsfxsize=0.30\hsize\epsfbox{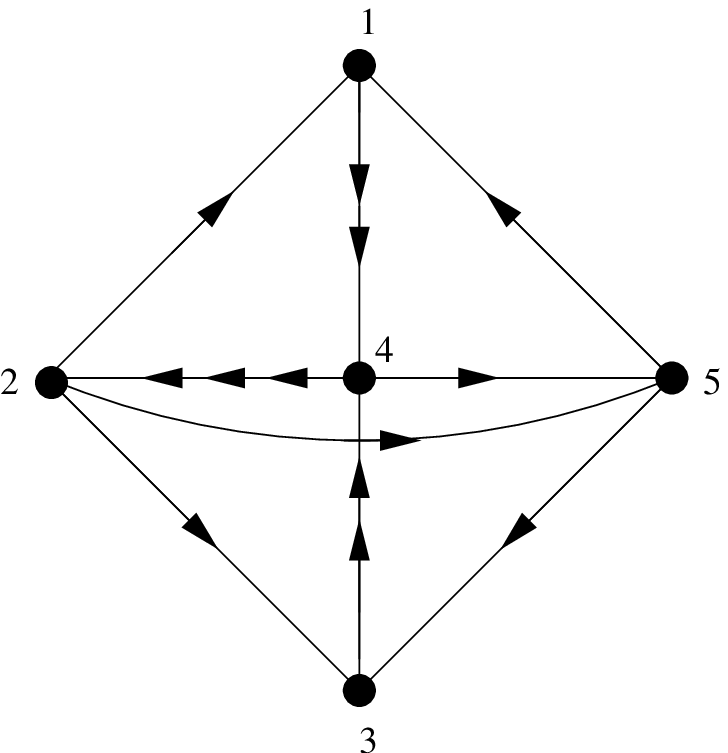}}
\centerline{\ninepoint\sl \baselineskip=8pt {\bf Figure 3:}
{\sl The $dP_2$ quiver resulting from Higgsing $X_{24}$.}}
\bigskip

It is easy to figure out the appropriate assignment of
divisors here: We simply take the divisors from our $dP_3$ model and
remove any $E_2$'s. In the case where bifundamental fields combine
and point in the same direction, the divisors never differ by 
more than an $E_2$, and thus there is no ambiguity. In the case
where bifundamentals combine that point in opposite directions, the
resulting divisor is the one corresponding to the bifundamental
that did not change direction, i.e. the one that had more flavors.

The $L_\alpha$ can be easily derived from our $dP_3$ example by simply
blowing down the divisor $E_2$. Since the $L_\alpha$ for the 
nodes from the $dP_3$ quiver that get combined are the same up to an
$E_2$, there is no ambiguity in how to assign the $L_\alpha$ for the 
$dP_2$ theory. We note that this also is true for any of the $E_i$ we could
have chosen to blow down. This yields (relabeling $E_3 \rightarrow E_2$)
\eqn\ldpii{\matrix{ {\rm Node} & L_\alpha &M_\alpha \cr
1 & E_1 &2N_2-N_1\cr
2 & 2D &2N_2\cr
3 & E_2 &N_1+N_2\cr
4 & D &N_2\cr
5 & 0&0.}}
\noindent 
The $M_\alpha$ in \ldpii\ give the gauge groups in the theory with added
wrapped D5's.  We take $J_1=E_1-E_2$ and $J_2=D-2E_1-E_2$, which 
satisfy $J_i\cdot c_1=0$ with $c_1 = 3D - E_1 - E_2$ for $dP_2$.
For the theory without wrapped branes, we get the following 
assignment of divisors and flavor charges:

\eqn\dpii{\matrix{{Q_{\alpha \beta}}  & \L & J_1 & J_2 & R \cr
X_{51} & E_1 & -1 & 2 & 2/7 \cr 
X_{53} & E_2 & 1 & 1 & 2/7 \cr
X_{25} & D - E_1 - E_2 & 0 & -2 & 2/7 \cr
X_{23} & D - E_1 & 1 & -1 & 4/7 \cr
X_{14} & D - E_1 & 1 & -1 & 4/7 \cr
X_{21} & D - E_2 & -1 & 0 & 4/7 \cr
X_{34} & D - E_2 & -1 & 0 & 4/7 \cr
X_{42} & D & 0 & 1 & 6/7 \cr
X_{45} & 2D - E_1 - E_2 & 0 & -1 & 8/7}}

\noindent Notice that the non-R flavor charges here are given
by linear combinations of the $U(1)$'s from $dP_3$ under which
$X_{42}$ is neutral, $J_1^{dP_2} = J_1^{dP_3} + J_3^{dP_3}$ and
$J_2^{dP_2} = J_2^{dP_3} - J_1^{dP_3}$. This is also consistent with 
the divisors assigned to these flavor charges, as one sees by
taking the appropriate linear combinations and removing any 
instances of the blown-down divisor.

The superpotential for this theory is \STD 
\eqn\wdpii{\eqalign{ W_{dP_2} &= X_{51}X_{14}X_{45} + 
X_{53}X_{34}X_{45} + X_{51}Y_{14}X_{42}X_{25} + 
X_{53}Y_{34}Y_{42}X_{25} \cr & + X_{21}X_{14}Y_{42} +
X_{23}X_{34}X_{42} + X_{21}Y_{14}Z_{42} + X_{23}Y_{34}Z_{42},}} where
we don't bother recording the exact coefficients.  This indeed obeys
the condition that every term has precisely one field with $\theta
_{\alpha
\beta}=1$.

The reader can easily verify that our 't Hooft anomaly conditions
are also satisfied: $\Tr R^3$ is indeed given by \delpth\ for
$n=2$, as required by \thooft.  The condition \amax\ of \IW\ is
indeed satisfied, showing that the geometry knows how to pick out
the correct superconformal $U(1)_R$, via $c_1$.  Finally, the flavor
't Hooft anomalies \fic\ vanish, as generally happens for
these string-constructed theories.    

We also check that our prescription for Seiberg duality
works. Dualizing on node 3 yields the other phase of $dP_2$, given in
Figure 4:

\bigskip
\centerline{\epsfxsize=0.30\hsize\epsfbox{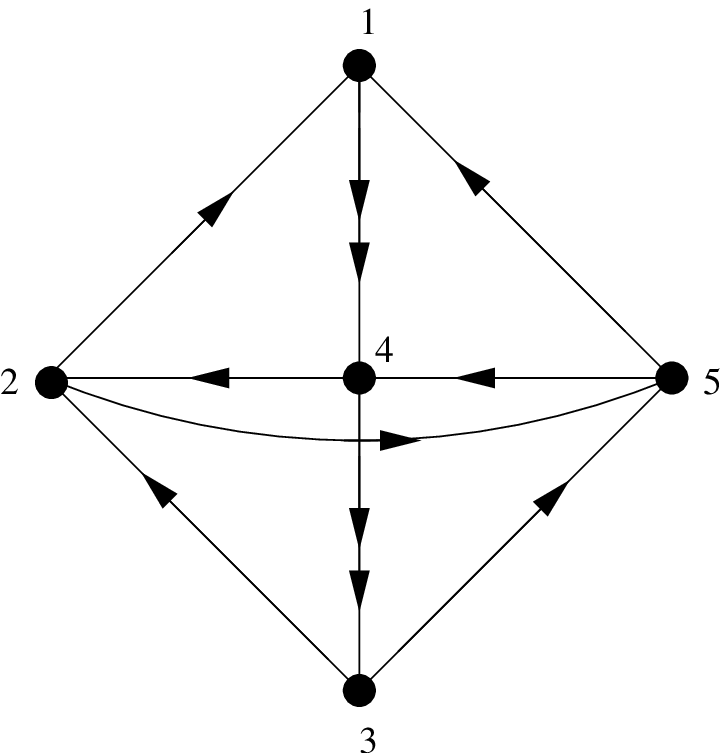}}
\centerline{\ninepoint\sl \baselineskip=8pt {\bf Figure 4:}
{\sl The other phase of $dP_2$.}}
\bigskip

\noindent The only $L$ that changes is $L_3$, which becomes
$L_3^\prime = L_2 + L_5 - L_3 = 2D - E_2$. It is easy to check
that this is consistent with the divisors one gets by 
appropriately Higgsing $dP_3$.

\subsec{Cone over $dP_1$}

It is useful here to Higgs the $dP_2$ theory to the $dP_1$ theory, since 
this is an especially simple example.
To obtain this theory, Higgs the field $X_{51}$ 
{}from $dP_2$, which corresponds to blowing down the exceptional curve $E_1$.
This yields the quiver in Figure 5.

\bigskip
\centerline{\epsfxsize=0.30\hsize\epsfbox{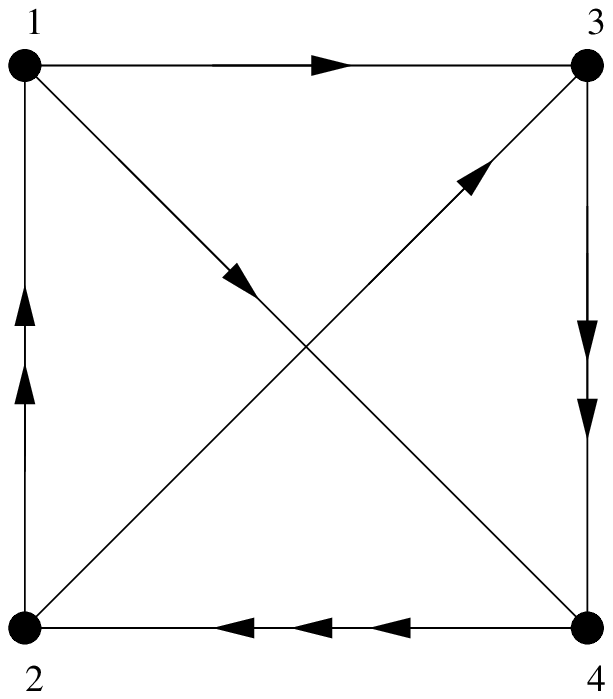}}
\centerline{\ninepoint\sl \baselineskip=8pt {\bf Figure 5:}
{\sl The $dP_1$ quiver}}
\bigskip

The $L_\alpha$, $D_\alpha$, and  $M_\alpha$ for wrapped branes are given by

\eqn\ddpiv{\matrix{ {\rm Node} & L_\alpha & D_\alpha & M_\alpha \cr
1 & 2D-E & E &-N_1\cr
2 & D & D &N_1\cr
3 & 2D & D-E &2N_1\cr
4 & 0 & 0& 0. }}

\noindent Note that here the flipped nodes are 1 and 3, where the $L_\alpha$ and
$D_\alpha$ differ. Here, $c_1 = 3D - E$, so
we take $J= D- 3E$. This yields the following fields and charges.

\eqn\dpi{\matrix{{Q_{\alpha \beta}}  & \L & J & R \cr
X_{13} & E & 3 & 1/4 \cr
X_{21} & D - E & -2  & 1/2 \cr
X_{34} & D - E & -2 & 1/2 \cr
X_{14} & D & 1 & 3/4 \cr
X_{23} & D & 1 & 3/4 \cr
X_{42} & D & 1 & 3/4 }}

The superpotential here is 
\eqn\wdpiis{W  = X_{42}X_{21}X_{14} 
+ X_{42}X_{23}X_{34} + X_{42}X_{21}X_{13}X_{34}}
which obeys the required conditions.

Seiberg dualizing on either node 1 or node 3 yields the same theory; one can 
check that the new $L_\alpha$ are identical to the original after 
relabeling nodes.

\subsec{Cone over $dP_4$}

As with the other del Pezzo surfaces, there are many different Seiberg
dual quiver theories possible for $dP_4$. Here, we will use the one
given in Figure 6 \refs{\FFHH,\IH}. It is straightforward to check
that by Higgsing $X_{67}$, one returns to the Model III $dP_3$ quiver.

\bigskip
\centerline{\epsfxsize=0.70\hsize\epsfbox{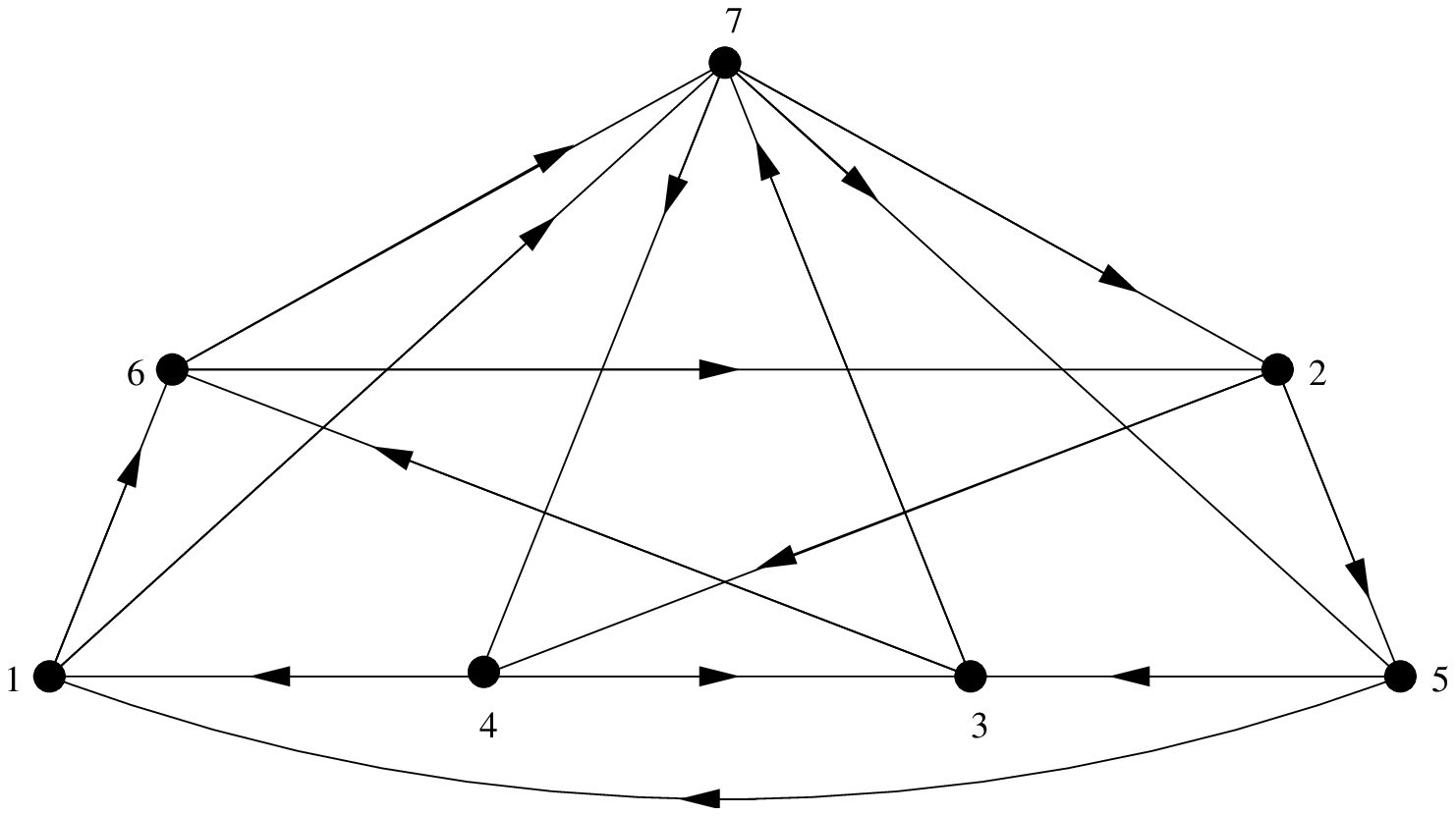}}
\centerline{\ninepoint\sl \baselineskip=8pt {\bf Figure 6:}
{\sl One possible quiver for $dP_4$}}
\bigskip

There is a unique assignment of $L_\alpha$ which reproduces the divisors
on the above $dP_3$ Model III theory.
These were found by enforcing that the field $X_{67}$
corresponds to the exceptional curve we're blowing down, $L_{67} = E_4$, 
and that the remaining divisors can only differ from their
$dP_3$ counterparts by this same exceptional curve.  We also list
the $M_\alpha = \sum _i N_i J_i \cdot L_\alpha$, relevant for the
theory with added wrapped D5s. 

\eqn\ldpiv{\matrix{ {\rm Node} & L_\alpha &M_\alpha \cr
1 & E_1 &N_4-N_1\cr
2 & 2D - E_2 - E_4 &N_4-N_1+N_2-N_3\cr
3 & E_3 &N_2-N_3+N_4\cr
4 & 2D - E_4 &2N_4-N_3\cr
5 & 0 &0\cr
6 & D - E_4 &N_4-N_3\cr
7 & D &N_4. \cr}}

On $dP_4$, the first Chern class is $c_1 = 3D - E_1 - E_2 - E_3 -
E_4$. Thus, we can take as our $J_i$ to be $J_1 = E_1 - E_2$, $J_2 =
E_2 - E_3$, $J_3 = E_3 - E_4$, and $J_4 = D - E_1 - E_2 - E_3$.
We thus find the divisors and charges to be

\eqn\dpIV{ \matrix{{Q_{\alpha \beta}} & \L & J_1 & J_2 & J_3 & J_4 & R \cr
X_{51} & E_1 & -1 & 0 & 0 & 1 & 2/5 \cr
X_{24} & E_2 & 1 & -1 & 0 & 1 & 2/5 \cr
X_{53} & E_3 & 0 & 1 & -1 & 1 & 2/5 \cr
X_{67} & E_4 & 0 & 0 & 1 & 0 & 2/5 \cr
X_{43} & D - E_1 - E_2 & 0 & 1 & 0 & -1 & 2/5 \cr
X_{25} & D - E_1 - E_3 & 1 & -1 & 1 & -1 & 2/5 \cr
X_{16} & D - E_1 - E_4 & 1 & 0 & -1 & 0 & 2/5 \cr
X_{41} & D - E_2 - E_3 & -1 & 0 & 1 & -1 & 2/5 \cr
X_{72} & D - E_2 - E_4 & -1 & 1 & -1 & 0 & 2/5 \cr
X_{36} & D - E_3 - E_4 & 0 & -1 & 0 & 0 & 2/5 \cr
X_{17} & D - E_1 & 1 & 0 & 0 & 0 & 4/5 \cr
X_{62} & D - E_2 & -1 & 1 & 0 & 0 & 4/5 \cr
X_{37} & D - E_3 & 0 & -1 & 1 & 0 & 4/5 \cr 
X_{74} & D - E_4 & 0 & 0 & -1 & 1 & 4/5 \cr
X_{75} & 2D - \sum_{i}E_i & 0 & 0 & 0 & -1 & 4/5 \cr}}
\bigskip

The superpotential for this theory \MW\ indeed obeys the condition
that each term has precisely one field with nonzero $\theta _{\alpha \beta}$.
(These charge and divisor assignments also apply for the $PdP_4$ 
case considered in \FFHH, which has a slightly different superpotential.)
We can also check that our 't Hooft anomaly conditions \delpth, \amax\ and
\fic\ are also satisfied.

It is also worth checking that one can Higgs this theory to $dP_3$ and
watch the divisor $E_4$ collapse in the same manner we observed in the
Higgsing of $dP_3$ down to $dP_2$. This indeed works; we note that
Higgsing $X_{67}$ and relabeling the node $6/7 \rightarrow 7$ produces
exactly the results found above.

Finally, we can immediately construct the quivers and $L_\alpha$ for
Seiberg dual theories.  For example, dualizing on node 2 yields
a quiver with  $L_2^{\prime} = L_7 + L_6 - L_2 = L_4 + L_5 - L_2 = E_2$
and all other $L_\alpha$ unchanged.

\centerline{\bf Acknowledgments}

We would like to thank M. Douglas, J. Kumar, J. Roberts, R.P. Thomas,
C. Vafa, and especially Mark Gross for discussions. We would also like
to thank Chris Herzog and James McKernan for alerting us to their
related work.   This work was supported by DOE-FG03-97ER40546.

\listrefs
\end